\documentclass{article}
 \usepackage{amsfonts}
\usepackage{amsmath}

\begin{document}

\title{\bf \large Tollmien-Shlichting and sound waves interaction: general description and nonlinear resonances}

\author{A. Perelomova, S.Leble,\\
 Technical University of Gda\'{n}sk, ul. Narutowicza 11/12\\
80-952 Gda\'{n}sk , Poland\\
 email:{leble@mif.pg.gda.pl} }


\date{March 25, 2002}


\begin{abstract}  The general hydro-thermodynamic system of
equations in 2+1 dimensions with arbitrary equations of state
(Taylor series approximation) is  split to eigen modes:
Tollmienn-Schlichting (TS) wave and two acoustic ones. A mode
definition is realized via local relation equations, extracted
from the linearization of the general system over boundary layer
flow. Each such connection defines the subspace and the
corresponding projector. So the division is performed locally and
could help in initial (or boundary) problems formulation. The
general nonlinearity account determines the specific form of
interaction between acoustic and vortical boundary layer
perturbation fields. After the next projecting to a subspace of
Orr-Sommerfeld equation solution for the TS wave and the
corresponding procedure for acoustics, the equations go to
one-dimensional system that describes evolution along the basic
stream. A new mechanism of nonlinear resonance excitation of the
TS wave by sound is proposed and modelled via four-wave
interaction. Subjectclass:  Primary 35Q30 ; Secondary  76F70 ,
35Q72; Keywords:
 keywords fluid mechanics, boundary layer, projector to eigen
modes, Tollmien-Schlichting waves, acoustic waves, nonlinear
resonance, N-wave system.
\end{abstract}

\maketitle

\section{Introduction}
 It is well-established fact now that a free-stream and a surface
disturbances strongly affect the processes in boundary layer which
compose a transition to a turbulent state. This transition in turn
determine such important parameters of a rigid body in fluid
mechanics as skin friction and heat transfer. The classification
of the free stream was done first by Kovasznay et al \cite{K}. It
is shown, that within linear approximation a general
small-amplitude unsteady disturbance in the free stream can be
decomposed into three independent different types: acoustic (A),
vortical and entropy modes. Only the first one  relates to
pressure fluctuations propagating with the sound speed, the last
two don't case any pressure perturbation. The common idea of many
investigations is to pick out length and time scales of each of
these disturbances that would make the possibility of the
Tollmienn-Schlichting (T-S) waves nonlinear generation \cite{Tam}.
This way the mechanism of T-S wave generation by convecting gusts
interacting with sound was justified and numerically investigated
\cite{Wu}. Appropriate scales of boundary roughness are proved to
generate T-S waves both theoretically and experimentally
\cite{rough}.

In spite of abundant efforts ( see the big introduction and the
citations in \cite{Wu}), devoted to the problem of a search of an
effective control mechanism that support a cumulative direct
energy exchange between T-S an acoustic modes. Let us mention
three important papers about the general, local and distributed
acoustic receptivity \cite{Ru}, \cite{Gold}, \cite{Cho}. The
results however do not look complete: we revisit the problem in
the all-perturbations approach \cite{L,Per}  starting  from
boundary layer (BL) as a background.  The perturbations are
considered only over the stationary boundary layer, we do not
account here (but plan to do it) the layer field as a dynamic
variable. This scheme gives a possibility to study mutual
interactions on the base of model integrable nonlinear evolution
equation. In our approach the description is local by the
construction and do not need averaging procedure \cite{Per1}. It
covers the known results and give new hopes for understanding
related phenomena appeared in papers from \cite{Koz} to \cite{KB}.
We would note that the initial stage of the process of its
structure changes manifests. This effect is similar to heating or
streaming generation by acoustic waves \cite{Mak}, it is  a
development of the stationary mode  and corresponds to an initial
stage of the BL reconstruction.

 In this paper we concentrate our efforts on the mathematical
formalism: introducing the complete set of basic modes we
transform the fundamental system of standard conservation laws of
fluid mechanics to a set of equivalent equations. In linear
approximation the specific choice of new dependent variables split
the system to the set of independent equations for the given
modes, the account of nonlinearity naturally introduce the
interaction by projecting the fundamental equations set in a
vector form. Going to the nonlinear description, we use iterations
inside the operator by the small parameter related to amplitude
(Mach number for acoustics) and viscosity (Reynolds number).  We
also analyze the possibilities of resonant interaction of
quasi-plane waves on the level of so-called N-wave systems
\cite{L}. Being integrable such systems admit explicit solutions
and plenty of conservation laws. Hence the detailed investigation
of situation is possible in this approximation.

The mentioned types of waves (T-S and A) are defined by
eigenvectors of the linearized system of dynamic conservation
equations for the free flow. Once defined, eigenvectors (or modes)
are fixed and independent on time. The process by which the free
stream disturbances are internalized to generate boundary-layer
instability waves is referred to as receptivity. The basic idea of
this paper is to define the T-S and acoustics modes as
eigenvectors of the same system for a viscous flow over a rigid
boundary. The eigenvectors of the viscous flow go to the known
limit in the free stream over a boundary. Our idea is to fix
relations between specific perturbations (velocity components,
density and pressure) for every mode followed from the linear
equations, to construct projectors and applied them for the
nonlinear dynamics investigation. Thus, in the linear dynamics the
overall field may be separated by projectors to the independent
modes at any moment. Fixing the relations when going to the
nonlinear flow, one goes to a system of coupled evolution
equations for the modes. These ideas have applied to nonlinear
dynamics of exponentially stratified gas and bubbly liquid
\cite{Per} and other problems \cite{L}.

The principal difference between this consideration and previous
ones is the necessity of variable coefficients incorporation that
arise from a boundary layer structure. It means that such
coordinate dependence impact the projectors structure: the
projector operator should be constructed by nonabelian entries. By
other words the matrix elements of the projector matrix will be
operator-valued ones. In fact we revisit and develop the first
attempt in this direction that had been made recently \cite{LTSA},
\cite{LPG}.

The T-S wave takes the place of a vortical mode in unbounded
space, the difference is due to the linearization on the different
background. When a stationary boundary-layer flow like Blausius
one appears, a linearization should be correctly proceeded with
account of the boundary-layer flow as a background, that would
lead obviously to the other features of the vortical mode then
that in unbounded space. An important feature of T-S mode is
non-zero disturbance of pressure already in the linear theory. In
the three-dimensional flow, there exist two T-S modes, two
acoustic ones (corrected by background flow), and one entropy mode
as well.

The equations of interaction are derived in Sec. 3 by means of the
division of the perturbation field on these subspaces and
projecting the system of the basic equations on the same
subspaces. This transformation that in fact is nothing but a
change of variables allows to proceed in a choice of adequate
approximation. Moreover we could  analyze possibilities of a
direct  nonlinear resonance interaction account. The results are
the following: two-wave and three-wave interaction does not
contribute due to the structure of the interaction terms in its
minimal possible order.  Hence (in this order) only the four-wave
interaction display a resonance structure. We derive the
correspondent four-wave equations  and analyze it in Sec.4.

\section{Basic equations treating the equations of state in the general
form.}

The mass, momentum and energy conservation equations read:
\begin{equation}\label{1}
  \frac{\partial \rho }{\partial t} +\vec{\nabla } (\rho \vec{v} )=0
\end{equation}

\begin{equation}\label{2}
  \rho \left[ \frac{\partial \vec{v} }{\partial t} +(\vec{v} \vec{\nabla } )%
\vec{v} \right] =-\vec{\nabla } p + \eta \Delta \vec{v} +\left( \varsigma +%
\frac{\eta }{3} \right) \vec{\nabla } (\vec{\nabla } \vec{v} )
\end{equation}

\begin{equation}\label{3}
  \rho \left[ \frac{\partial e}{\partial t} +(\vec{v} \vec{\nabla } )e\right]
+p\vec{\nabla } \vec{v} = \chi \Delta T + \varsigma \left(
\vec{\nabla } \vec{v} \right)^{2}  +\frac{\eta }{2} \left(
\frac{\partial v_{i} }{\partial x_{k} }
+\frac{\partial v_{k} }{\partial x_{i} } -\frac{2}{3} \delta _{ik} \frac{%
\partial v_{l} }{\partial x_{l} } \right) ^{2}
\end{equation}

Here, $\rho, p$ are density and pressure, $e, T$ - internal energy
per unit mass and temperature, $\eta, \varsigma,$ are shear, bulk
viscosities, and $\chi$ -  thermal conductivity coefficient
respectively (all supposed to be constants), $\vec{v} $ is a
velocity vector, $x_{i} $  - space coordinates. Except of the
dynamical equations (1,2,3), the two thermodynamic relations are
necessary: $e(p,\rho ),T(p,\rho )$ . To treat a wide variety of
substances, let us use the general form of the caloric equation
(energy) as expansion in the Taylor  series:
\begin{equation}\label{4}
  \rho _{0}e=E_{1}p+\frac{E_{2}p_{0}}{\rho _{0}}\rho +\frac{E_{3}}{p_{0}}%
p^{2}+\frac{E_{4}p_{0}}{\rho _{0}^{2}}\rho ^{2}+\frac{E_{5}}{\rho _{0}^{{}}}%
p\rho +\frac{E_{6}}{p_{0}\rho _{0}^{{}}}p^{2}\rho +\frac{E_{7}}{\rho _{0}^{2}%
}p\rho ^{2}+\frac{E_{8}}{p_{0}^{2}}p^{3}+\frac{E_{9}p_{0}}{\rho _{0}^{2}}%
p^{3} +\ldots,
\end{equation}
and the thermic one
\begin{equation}\label{5}
  T =\frac{\Theta _{1} }{\rho _{0} C_{v} } p +\frac{\Theta
_{2} p_{0} }{\rho _{0}^{2} C_{v} } \rho   + \ldots.
\end{equation}
The background values for unperturbed medium are marked by zero,
perturbations of pressure and density are denoted by the same
characters (no confusion is possible since only perturbations
appear below ), $C_{v} $ means the specific heat per unit mass at
constant volume, $E_{1} ,...\Theta _{1} ,...$ are dimensionless
coefficients. The two-dimensional flow in the coordinates x
(streamwise distance from the plate (model) leading edge), z
(wall-normal distance from a model surface) relates to the
two-component velocity vector
\begin{equation}\label{6}
\vec{v} =\left( u,w \right) +\vec{u} _{0},
\end{equation}
where $\vec{u} _{0}  =(U_{0} (z),0) $  denotes the background
streamwise velocity and $\left( u,w\right) $  stands for velocity
perturbations.
The system (\ref{1} - \ref{3})  with account of (\ref{4}, \ref{5},
\ref{6} ) yields in
\begin{equation}\label{8}
 \begin{array}{c}
\rho _{0} \left( \frac{\partial u}{\partial t} +U_{0} \frac{\partial u}{%
\partial x} +w\frac{\partial U_{0} }{\partial z} \right)
+\frac{\partial p}{\partial x} -\eta \Delta u - \\
\left(\varsigma + \frac{\eta }{3}
\right) \left( \frac{\partial ^{2} u}{\partial x^{2} } +\frac{\partial ^{2} w%
}{\partial x\partial z} \right) = -\rho _{0} u\frac{\partial
u}{\partial x} -\rho _{0} w\frac{\partial u}{\partial z}
+\frac{\rho}{\rho _{0} } \frac{\partial p}{\partial x}
\\
\rho _{0} \left( \frac{\partial w}{\partial t} +U_{0} \frac{\partial w}{%
\partial x} \right) +\frac{\partial p}{\partial z} -\eta \Delta
w-\left( \varsigma +\frac{\eta }{3} \right) \left( \frac{\partial ^{2} u}{%
\partial x\partial z} +\frac{\partial ^{2} w}{\partial z^{2} } \right)
= \\
-\rho _{0} w\frac{\partial w}{\partial z} -\rho _{0} u\frac{\partial w}{%
\partial x} +\frac{\rho ^{\prime}}{\rho _{0} } \frac{\partial p}{\partial z}
\\
\frac{\partial p}{\partial t} +U_{0} \frac{\partial p}{%
\partial x} +c^{2} \rho _{0} \left( \frac{\partial u}{\partial x}
+
\frac{%
\partial w}{\partial z} \right) = \\
\frac{\chi }{E_{1} } \left( \frac{\Theta _{1} }{\rho _{0} C_{v} }
\Delta p+\frac{\Theta _{2} p_{0} }{\rho _{0}^{2} C_{v} } \Delta
\rho  \right) \left[ pZ +\rho c^{2}S \right] -u\frac{\partial
p}{\partial x} -w\frac{\partial p}{\partial z},
\end{array}
\end{equation}
where the constants $Z$ and $S$ are defined by
\begin{equation}\label{T,S}
\begin{array}{c}
  Z =  \left( -1+2\frac{1-E_{2} }{E_{1} } E_{3}
+E_{5} \right)/ E_{1} , \\
 S  = \frac{1}{1-E_{2} } \left( 1+E_{2} +2E_{4} +\frac{1-E_{2}
}{E_{1} } E_{5} \right).
\end{array}
\end{equation}
The constant $c=\sqrt{\frac{p_{0} (1-E_{2} )}{\rho _{0} E_{1} } }
$ has the sense of linear sound velocity in the medium under
consideration when $U_0 = 0 $. The right-hand side of equations
involves the quadratic nonlinear and viscous terms as well as
linear ones related to thermal conductivity, no cross
viscous-nonlinear terms accounted. In fact, the third equation
follows from the energy balance and continuity equation. No
assumptions on flow compressibility was not done yet.

\section{Modes  in the linear approximation.}

The basic system (\ref{1}-\ref{3}) contains four dynamic equations
and therefore, there are four independent modes of the linear
flow: two acoustic ones, vorticity and heat modes. This is a
classification by Kovasznay who defined the acoustic modes as
isentropic and irrotational flow,  and refers to the two last ones
as to frozen motions, the vorticity one relating to the absence of
pressure and density perturbations and the heat one relating to
the very density perturbation. The system (\ref{1}-\ref{3})
involves three equations indeed, therefore only the three modes
may be extracted - two acoustic and vorticity ones. That is due to
the structure of the heat mode where the only density perturbation
occurs. Strictly speaking, the term treating a thermal
conductivity in the third of equations (\ref{3}) includes the
density perturbations and the linearized system (\ref{8}) (that
defines the modes as possible types of flow) is not closed. If
there was no this term at all there is a simple linear relation
between density and pressure perturbations for the both acoustic
modes. The presence of thermal conductivity corrects this relation
as it was shown in \cite{RS}. When the effects of thermal
conductivity are small, the corresponding terms may be placed to
the right-hand side of equation together with nonlinear ones and
be accounted further. So the excluding of the dynamic equation for
density serve just a simplification of a problem suitable in the
view of its extraordinary complexity when the heat mode is out of
the area of interest.

The main idea is to define modes accordingly to the specific
relations of the basic perturbation variables following from the
linearized system of dynamic equations (\ref{1}-\ref{3}). In
general, the procedure is algorithmic and may be expressed as
consequent steps : to find the dispersion relation and its roots
that determine all possible modes, and to define relations between
specific variables for every mode. Later, projectors follow
immediately from these relations , they separate every mode from
the overall field of the linear flow exactly and serve as a tool
for the nonlinear dynamics investigation.

Using the left-hand  part of the system (\ref{8}) as a basis of
modes definition and introducing the two non-dimensional functions
$V_{0} (z)=U_{0} (z)/U_{\infty } $ , $\phi(z)=V_{0z} (z)l_{0} $ ,
we rewrite it in the non-dimensional variables
\begin{equation}\label{dim}
  x_{*} =x/l_{0} , w_{*} =w/U_{\infty } , u_{*}
=u/U_{\infty } , t_{*} =tU_{\infty } /l_{0} , p_{*} =p/\rho _{0}
U_{\infty }^{2}.
\end{equation}
The value $U_{\infty } $  marks velocity of a flow far from the boundary, and $%
l_{0} $ - boundary layer width.
In the new variables (asterisks will be later omitted) (\ref{8})
with zero right side reads :
\begin{equation}\label{71}
  \frac{\partial p}{\partial t} +V_{0} \frac{\partial p}{\partial x}
+\epsilon ^{-2} \left( \frac{\partial u}{\partial x} +\frac{\partial w}{%
\partial z} \right) =0
\end{equation}
\begin{equation}\label{72}
  \frac{\partial u}{\partial t} +V_{0} \frac{\partial u}{\partial x} +\phi w+%
\frac{\partial p}{\partial x} - {Re} ^{-1} \Delta u-R^{-1} \left(
\frac{\partial ^{2} u}{\partial x^{2} } +\frac{\partial ^{2}
w}{\partial x\partial z} \right) =0
\end{equation}

\begin{equation}\label{73}
  \frac{\partial w}{\partial t} +V_{0} \frac{\partial w}{\partial x} +\frac{%
\partial p}{\partial z} - {Re} ^{-1} \Delta w-R^{-1} \left( \frac{%
\partial ^{2} u}{\partial x \partial z} +\frac{\partial ^{2} w}{\partial
z^{2} } \right) =0
\end{equation}
with parameters $\epsilon =U_{\infty } /c$  (the Mach number), $
{Re} =U_{\infty } l_{0} \rho _{0} /\eta $  is the Reynolds number
base on the length scale, and $R=U_{\infty } l_{0} \rho _{0}
/\left( \eta /3+\varsigma \right) $ .

\subsection{ The Tollmienn-Schlichting mode.}

Formally, the limit of incompressible fluid ( $\epsilon =0$ )
corresponds to the vorticity mode. The first relation for velocity
components is well-known \cite{Wu}:
\begin{equation}\label{div}
  \partial u/\partial x+ \partial w/\partial z=0.
\end{equation}

From the equations(\ref{72},\ref{73}) an expression for the
pressure perturbation follows:
\begin{equation}\label{1rel}
  2\phi \partial w/\partial x+\Delta p=0
\end{equation}
where $\Delta =\partial ^{2} /\partial z^{2} - k^{2} $ stands for
the  Laplacian so far the $%
\partial /\partial x$ equivalent operator (multiplier) $-ik$ is used.
 Both (\ref{div}) and  (\ref{1rel}) define the TS mode due to
relations of the specific perturbations of pressure and velocity
components. Since the geometry of the viscous flow over boundary
supposes strong non-uniformity in the vertical direction, all
disturbances may be thought in the basis of not plane waves but in
the functions like $\Psi(x,z)=\psi (z)\exp (i\omega t-ikx)$ . So,
one has to leave vertical derivatives that usually are large in
comparison with the horizontal ones. The result is rather obvious
hence we do not introduce a special small parameter. The vector of
the T-S mode may be chosen as:
\begin{equation}\label{TS}
  T=\left(
\begin{array}{c}
p_{TS} \\
u_{TS} \\
w_{TS}%
\end{array}
\right) =\left(
\begin{array}{c}
1 \\
-\frac{1}{2k^{2} } \frac{\partial }{\partial z} \frac{1}{\phi } \Delta \\
\frac{1}{2ik\phi } \Delta%
\end{array}
\right) p_{TS}
\end{equation}
Note also, that the equations (\ref{71}-\ref{73}) yield the
well-known equation for the TS mode \cite{Sch}, when rewritten for
the new variable such the stream function ( $u=\partial \Psi
/\partial z$ , $w=-\partial \Psi /\partial x$, with the obvious
restriction to the solenoidal velocity field (\ref{div}):
\begin{equation}\label{psi}
  \Delta \partial \Psi /\partial t+V_{0} \Delta \partial \Psi /\partial
x-\partial \Psi /\partial x\cdot \partial \phi /\partial z- {Re}
^{-1} \Delta ^{2} \Psi =0
\end{equation}
The well-known Orr-Sommerfeld (OS) equation follows from
(\ref{psi}):
\begin{equation}\label{12}
  \left[ V_{0} (z)-c\right] \left[ \partial ^{2} \psi /\partial z^{2} -k^{2}
\psi \right] -\psi \partial \phi /\partial z=\frac{i}{{Re} k}
\left[
\partial ^{4} \psi /\partial z^{4} -2k^{2} \partial ^{2} \psi /\partial
z^{2} +k^{4} \psi \right]
\end{equation}

That equation is an initial point of the laminar flow stability
theory and for every pair (k,Re) determines an eigenfunction $\psi
(z)$ and complex phase velocity $c=\omega /k=c_{r} +ic_{i} $ . The
sign of $c_{i} $ is namely a criterion of flow stability: a
negative value corresponds to the growth of perturbation and
therefore to the non-stability of the flow.

\subsection{Acoustic modes}

The potential flow imposed two acoustic modes with
 $$\partial u/\partial
z-\partial w/\partial x=0. $$
 In the limit of $ {Re}^{-1} =0$ , $%
R_{}^{-1} =0$ , $\phi =0$ , (\ref{71}-\ref{73}) naturally goes to
the acoustic modes dispersion relation that is directly connected
with the wave operator.
\begin{equation}\label{13}
  \epsilon ^{2} \frac{\partial ^{2} }{\partial t^{2} } -\Delta
\end{equation}
We would not consider here the perturbation velocity field
changes, forced by the ambient movement of the fluid. It could be
account by the perturbation theory to be developed here. Then, two
acoustic modes are defined with relations between specific
perturbations:

\begin{equation}\label{14}
 A_{1} =\left(
\begin{array}{c}
p_{A1} \\
u_{A1} \\
w_{A1}%
\end{array}
\right) =\left(
\begin{array}{c}
1 \\
-i\epsilon k\Delta ^{-1/2} \\
\epsilon \frac{\partial }{\partial z} \Delta ^{-1/2}%
\end{array}
\right)p_{A1}, A_{2} =\left(
\begin{array}{c}
p_{A2} \\
u_{A2} \\
w_{A2}%
\end{array}
\right) =\left(
\begin{array}{c}
1 \\
i\epsilon k\Delta ^{-1/2} \\
-\epsilon \frac{\partial }{\partial z} \Delta ^{-1/2}%
\end{array}
\right)p_{A2}
\end{equation}

Here the square root of the operator $\Delta$ is defined as
integral operator via Fourier transform.

\section{ Projectors in the linear problem.}

The TS and two acoustic modes are determined by relations of
specific perturbations or, in the vector form (\ref{TS}),
(\ref{14}). The superposition of such disturbances appear as all
possible types of flow except of the heat mode. Every mode is
completely defined by one of specific perturbations - pressure or
velocity components since there are strict and local relations
between them. Practically, in a linear flow, the overall
perturbation may be de-coupled into modes by the corresponding
orthogonal projectors. The  arbitrary perturbation then is a sum
of modes which, taking into account (\ref{TS}), (\ref{14}) looks:
\begin{equation}\label{link}
 \left(
\begin{array}{c}
p \\
u \\
w%
\end{array}
\right) =\left(
\begin{array}{c}
p_{A1} +p_{A2} +p_{TS} \\
u_{A1} +u_{A2} +u_{TS} \\
w_{A1} +w_{A1} +w_{TS}%
\end{array}
\right) =\left(
\begin{array}{c}
p_{A1} +p_{A2} +p_{TS} \\
Hp_{A1} -Hp_{A2} +Kp_{TS} \\
Mp_{A1} -Mp_{A2} +Qp_{TS}%
\end{array}
\right)
\end{equation}
with operators
\begin{equation}\label{sqrDelta}
\begin{array}{c}
 H=-\epsilon \Delta ^{-1/2} ik,\\ M=\epsilon \Delta ^{-1/2} \partial
/\partial z ,\\ K=-\frac{1}{2k^{2} } \partial /\partial
z\frac{1}{\phi } \Delta  , \\Q=\frac{1}{2\phi ik} \Delta
\end{array}
\end{equation}

The link (\ref{link}) may be considered as a one-to-one map of
dynamical variables
that immediately yields in the case of TS wave the
projector :
\begin{equation}\label{pTS}
  P_{TS} =
  \left(
\begin{array}{c}
0 \\
0 \\
0%
\end{array}
\begin{array}{c}
\\
\end{array}
\begin{array}{c}
\\
\end{array}
\begin{array}{c}
-2k^{2} \Delta ^{-1} \phi \partial /\partial z\Delta ^{-1} \\
\partial ^{2} /\partial z^{2} \Delta ^{-1} \\
ik\partial /\partial z\Delta ^{-1}%
\end{array}
\begin{array}{c}
-2ik^{3} \Delta ^{-1} \phi \Delta ^{-1} \\
ik\partial /\partial z\Delta ^{-1} \\
-k^{2} \Delta ^{-1}.%
\end{array}
\right)
\end{equation}
For a right and left acoustic waves one has
$$
P_{A1} =
$$
\begin{equation}\label{pA1}
 \frac{ \Delta^{1/2}}{2} \left(
\begin{array}{ccc}
1 \\
\frac{-ik}{\epsilon}   \\
\frac{1}{ \epsilon}   \frac{\partial}{\partial z}
\end{array}
\begin{array}{c}
2k^{2} \Delta ^{-1/2} \phi \frac{\partial}{\partial z}\Delta ^{-1}
-\frac{ik}{\epsilon} \Delta ^{-1/2} \\
-ik^{2}\epsilon \left[ 2k \Delta ^{-1} \phi
\frac{\partial}{\partial z}
\Delta ^{-1} -\frac{i}{ \epsilon} \Delta^{-1/2} \right] \\
k\epsilon  \frac{\partial}{\partial z}  \left[2 k \Delta ^{-1}
\phi \frac{\partial}{\partial z} \Delta ^{-1}- \frac{
ik}{\epsilon} \Delta ^{-1/2} \right]
\end{array}
\begin{array}{c}
2ik^{3} \Delta ^{-1/2} \phi \Delta ^{-1} +\frac{1}{ \epsilon}  \frac{\partial}{\partial z}  \\
-ik\epsilon   \left[ 2ik^{3} \Delta ^{-1} \phi \Delta ^{-1}
+\frac{1}{ \epsilon} \Delta^{-1/2} \frac{\partial}{\partial z} \right] \\
\epsilon   \frac{\partial}{\partial z} \left[ 2ik^{3} \Delta ^{-1}
\phi \Delta ^{-1} +\frac{1}{ \epsilon} \Delta^{-1/2}
 \frac{\partial}{\partial z} \right]
\end{array} \right)
\end{equation}

  $$P_{A2}=$$
\begin{equation}\label{pA2}
   \frac{\Delta^{1/2} }{2}\left(
\begin{array}{ccc}
 1 \\
\frac{ik}{\epsilon} \\
-\frac{1}{\epsilon} \frac{\partial}{\partial z}
\end{array}
\begin{array}{c}
2k^{2} \Delta ^{-1/2} \phi \frac{\partial}{\partial z} \Delta
^{-1}
+\frac{ik}{\epsilon}   \\
 2ik^{2} \epsilon  \left[ k \Delta ^{-1} \phi
\frac{\partial}{\partial z}
z\Delta ^{-1} +\frac{i }{\epsilon } \Delta ^{-1/2} \right] \\
-2k\epsilon \frac{\partial}{\partial z} \left[ \Delta^{-1} \phi
\frac{\partial}{\partial z}\Delta ^{-1} +\frac{i}{\epsilon} \Delta ^{-1/2} \right]%
\end{array}
\begin{array}{c}
2ik^{3} \Delta ^{-1/2} \phi \Delta ^{-1} -\frac{1}{\epsilon}
\frac{\partial}{\partial z} \\
2ik\epsilon   \left[ ik^{3} \Delta ^{-1} \phi \Delta ^{-1}
-\frac{1}{\epsilon} \Delta ^{-1/2} \frac{\partial}{\partial z}\right] \\
-2\epsilon  \frac{\partial}{\partial z}\left[ ik^{3} \Delta ^{-1}
\phi
\Delta ^{-1} -\frac{1}{\epsilon} \Delta ^{-1/2} \frac{\partial}{\partial z}\right]%
\end{array}
\right)
\end{equation}

The projectors possess all properties of orthogonal projectors and
their sum is unit matrix since all eigenvectors of linear system
are accounted.
$$P_{TS}^2 = P_{TS}, ..., P_{TS} +P_{A1} +P_{A2} =I,P_{TS} \cdot
P_{A1} =P_{TS} \cdot P_{A2} =...=0, $$ where $I$ and 0 are unit
and zero matrices correspondingly. We would stress that the
operators defined by the (\ref{pTS}, \ref{pA1}, \ref{pA2}) contain
operator-valued matrix elements. Hence the derivation of their
explicit form as well as the properties mentioned above could be
checked taking the nonabelian nature of the operators into
account. In the linear flow, the projectors separate every mode
from the overall perturbation, for example:
\begin{equation}\label{ts1}
  P_{TS} \left(
\begin{array}{c}
p \\
u \\
w%
\end{array}
\right) =\left(
\begin{array}{c}
p_{TS} \\
u_{TS} \\
w_{TS}%
\end{array}
\right)
\end{equation}
 and so on. Moreover, acting by a projector on the basic system of
dynamic equations (\ref{71}-\ref{73}), yields a linear evolution
equation for the mode assigned to this projector. One produces
three equations indeed for every specific perturbations which are
essentially the same with account of relations (\ref{1rel}),
(\ref{14}). For the first rightwards progressive acoustic mode the
evolution equation reads:
\begin{equation}\label{A1}
  \partial p_{A1} /\partial t+V_{0} \partial p_{A1} /\partial x+\epsilon
^{-1} \Delta ^{1/2} p_{A1} =0
\end{equation}
The equation for the second (opposite directed) acoustic mode is
produced by acting the projector $p_{A2}$ on the basic system and
differ from {\ref{A1}} only by the sign before the last term.
Combining equations for these directed acoustic modes one arrives
at the wave equation of second order
$$\left( \partial /\partial t+V_{0} \partial /\partial x\right)
^{2} p-\epsilon ^{-2} \Delta p=0.$$
 This equation appears as a
limit  of more general one: that relates to both acoustic modes
and can be found in \cite{B}.

\section{Nonlinear flow: coupled dynamic equations.}

In the  dimensionless variables introduced by (\ref{6}), the
dynamic equations with account of nonlinear terms of the second
order, look:
\begin{equation}\label{NE}
\begin{array}{c}
  \frac{\partial p}{\partial t} +V_{0} \frac{\partial p}{\partial x}
+\epsilon ^{-2} \left( \frac{\partial u}{\partial x} +\frac{\partial w}{%
\partial z} \right) =\tilde{\varphi } _{1}\\
 \frac{\partial u}{\partial t} +V_{0} \frac{\partial u}{\partial x} +\phi w+%
\frac{\partial p}{\partial x} -  {Re} ^{-1} \Delta u-R^{-1} \left(
\frac{\partial ^{2} u}{\partial x^{2} } +\frac{\partial
^{2} w}{\partial x\partial z} \right) =\tilde{\varphi } _{2}\\
\frac{\partial w}{\partial t} +V_{0} \frac{\partial w}{\partial x} +\frac{%
\partial p}{\partial z} - {Re} ^{-1} \Delta w-R^{-1} \left( \frac{%
\partial ^{2} u}{\partial x\partial z} +\frac{\partial ^{2} w}{\partial
z^{2} } \right) =\tilde{\varphi} _{3}
\end{array}
\end{equation}
with a vector of the second-order nonlinear terms $\tilde{\psi } $
in the right-hand side (a non-dimensional value $\rho _{*} =\rho
/\rho _{0} $  is used in the right-hand side, the asterisk will be
omitted later) :
\begin{equation}\label{34}
   \widetilde{ \varphi  } =\left(
\begin{array}{c}
\tilde{\varphi } _{1} \\
\tilde{\varphi } _{2} \\
\tilde{\varphi } _{3}%
\end{array}
\right) = \left(
\begin{array}{c}
-u\frac{\partial p}{\partial x} -w\frac{\partial p}{\partial z}
+\left(
\frac{\partial u}{\partial x} +\frac{\partial w}{\partial z} \right) \left[%
\ Z\ p\ +  \epsilon ^{-2}\ S \rho   \right] \\
-u\frac{\partial u}{\partial x} -w\frac{\partial u}{\partial z} +\rho \frac{%
\partial p}{\partial x} \\
-u\frac{\partial w}{\partial x} -w\frac{\partial w}{\partial z} +\rho \frac{%
\partial p}{\partial z}%
\end{array}
\right)
\end{equation}

One could rewrite a system (\ref{NE}) in another way:
\begin{equation}\label{NE1}
\frac{\partial }{\partial t} \varphi +L\varphi =\tilde{\varphi },
\end{equation}
where the vector state  $\varphi$ is a specific perturbations
column (vector of the fluid state)
\begin{equation}\label{NE1}
\varphi =\left(
\begin{array}{c}
p \\
u \\
w%
\end{array}
\right)
\end{equation}
  and L is a matrix operator
\begin{equation}\label{L}
\left(
\begin{array}{ccc}
V_{0} \partial /\partial x & \epsilon ^{-2} \partial /\partial x &
\epsilon
^{-2} \partial /\partial z \\
\partial /\partial x & V_{0} \partial /\partial x- {Re} ^{-1} \Delta
-R^{-1} \partial ^{2} /\partial x^{2} & \phi -R^{-1} \partial ^{2}
/\partial
x\partial z \\
\partial /\partial z & -R^{-1} \partial ^{2} /\partial x\partial z & V_{0}
\partial /\partial x- {Re} ^{-1} \Delta -R^{-1} \partial ^{2}
/\partial z^{2}%
\end{array}
\right)
\end{equation}

All the projectors do commute with operators $\partial /\partial
t\cdot I$ and L, so one can act by projectors on the system of
equations (\ref{NE1}), (\ref{L}) directly thus obtaining the
evolution equation for the correspondent mode. There are three
such equations for the specific perturbations p, u, w for every
mode (that are equivalent) accounting the relations between these
specific perturbations. So, as independent variable for every mode
a single specific perturbation, such as pressure or one velocity
component, may be chosen.

Due to the existing tradition it is convenient to use the stream
function as a basic for the TS mode and pressure perturbations for
the acoustic modes.

Acting by the $P_{TS} $  on the both sides of (\ref{NE1}), one
gets an evolution equation:
\begin{equation}\label{NTS}
\begin{array}{c}
  \Delta \partial \Psi /\partial t+V_{0} \Delta \partial \Psi /\partial
x-\partial \Psi /\partial x\cdot \partial \phi /\partial z - {Re}
^{-1} \Delta ^{2} \Psi =\frac{\partial }{\partial z} [ -u \partial u/\partial x \\
- w\partial u/\partial z+\rho \partial p/\partial x ] -\frac{%
\partial }{\partial x} \left[ -u\partial w/\partial x-w\partial w/\partial
z+\rho \partial p/\partial z \right].
 \end{array}
\end{equation}

 Next it should be noted that in the right-hand nonlinear side p, u, w
are overall perturbations to be presented as a sum of specific
perturbations of all modes:
\begin{equation}\label{tr}
  \begin{array}{c}
  u=\partial \Psi /\partial z+\epsilon \Delta ^{-1/2} \partial p_{A1}
/\partial x-\epsilon \Delta ^{-1/2} \partial p_{A2} /\partial x \\
  w=-\partial \Psi /\partial x+\epsilon \Delta ^{-1/2} \partial p_{A1}
/\partial z-\epsilon \Delta ^{-1/2} \partial p_{A2} /\partial z \\
  p=2\partial ^{2} /\partial x^{2} \Delta ^{-1} (\phi \Psi )+p_{A1} +p_{A2}
\end{array}
\end{equation}

Indeed, there is also a density perturbation in the right-hand
nonlinear side that was not involved in the left-hand linear one
at all. The continuity equation reads
\begin{equation} \label{ce}
  \partial \rho /\partial t+V_{0} \partial \rho /\partial x+(\partial
u/\partial x+\partial w/\partial z)= -\partial (\rho u)/\partial
x-\partial  ( \rho  w)/ \partial z
\end{equation}

Comparing linear left-hand side of equation (\ref{ce}) with that
of the first equation from (\ref{A1}), the obvious relations for
the both acoustic modes follow:

$\rho _{A1} =\epsilon ^{2} p_{A1} $ , $\rho _{A2} =\epsilon ^{2} p_{A2} $ .

A limit $\epsilon =0$  yields in the TS mode for incompressible flow: $\rho
_{TS} =0$ . Therefore, the last relation for the overall density
perturbation looks
\begin{equation}\label{ro}
  \rho =\epsilon ^{2} p_{A1} +\epsilon ^{2} p_{A2}
\end{equation}
 Finally, (\ref{NTS}) goes to
\begin{equation}\label{NTS1}
\begin{array}{c}
\Delta  \Psi_{t} + V_{0} \Delta \Psi_{x} - \Psi _{x}  \phi _{z} -
{Re} ^{-1} \Delta^{2} \Psi =- \Psi _{z} \Delta \Psi _{x}
+\Psi _{x} \Delta \Psi _{z} - \\
 \epsilon ( \Delta \Psi \Delta^{1/2} (p_{A1} - p_{A2} )+ \Delta
\Psi_{x}  \Delta^{-1/2} (p_{A1} -p_{A2} )_{x} + \\
\Delta \Psi _{z}  \Delta^{-1/2} (p_{A1} -p_{A2} )_{z}  ) +
O(\epsilon ^{2} ).
\end{array}
\end{equation}
Derivatives are marked with lower indices. The first two nonlinear
terms in the right-hand  side of (\ref{NTS1}) expresses the TS
mode self-action, and the last ones - cross acoustic-vorticity
terms responsible for the acoustic mode influence on the TS mode
propagation. The structure of the quadratic nonlinear column
(\ref{NE1}) yields in the absence of quadratic acoustic terms in
(\ref{NTS1}).

In the limit of $V_{0} =0$ , $\phi =0$  the only self-action give
the well-known evolution equation for vorticites transition
\cite{Sch} follows from (\ref{NTS1}):
$$\Delta \Psi _{t} - {Re} ^{-1} \Delta ^{2} \Psi +\Psi _{z}
\Delta \Psi _{x} -\Psi _{x} \Delta \Psi _{z} =0.$$

Let  an acoustic field consists of only the first mode. Acting by projector $%
P_{A1} $ on the system (\ref{34}), we get an evolution equation
for this mode:
\begin{equation}\label{a0}
\begin{array}{c}
  \partial p_{A1} /\partial t + V_{0} \partial p_{A1} /\partial x+\Delta ^{1/2}
p_{A1} /\epsilon = \frac{1}{2}[ -u\partial p/\partial
x-w\partial p/\partial z+ \\
\left(
\partial u/\partial x+\partial w/\partial z\right) \left( Zp+S\rho
\epsilon ^{-2} \right) ] +\\
   (-\partial ^{2} /\partial x^{2} \Delta ^{-1} \phi \partial /\partial
z\Delta ^{-1} +\partial /\partial x(1/2\epsilon )\Delta ^{-1/2}
)\left[
- u\partial u/\partial x-w\partial u/\partial z+\rho \partial p/\partial x%
\right] + \\
  (\partial ^{3} /\partial x^{3} \Delta ^{-1} \phi \Delta ^{-1} +(1/2\epsilon
)\Delta ^{-1/2} \partial /\partial z)\left[ -u\partial w/\partial
x-w\partial w/\partial z+\rho \partial p/\partial z\right]. \\
\end{array}
\end{equation}

Here constants Z and S are defined earlier by (\ref{T,S}). The
variables p, u, w, are overall perturbations accordingly to
(\ref{tr}), with $p_{A2}=0 $ . So, (\ref{a0}) goes to  the final
version for the directed acoustic mode
\begin{equation}\label{a}
   \ \begin{array}{c}
       \epsilon \left( \partial p_{A1} /\partial t+V_{0} \partial p_{A1} /\partial
x\right) +\Delta ^{1/2} p_{A1} =\\
   \frac{\epsilon }{2} [ -\Psi _{z} p_{A1x} +\Psi _{x} p_{A1z}] +
\epsilon \Delta ^{-1/2}  [ -\Psi _{z} \Delta ^{1/2} p_{A1x}
+ \\
\Psi _{x} \Delta ^{1/2} p_{A1z} -2\Psi _{xz} \Delta ^{-1/2}
p_{A1xx} -2\Psi _{zz} \Delta ^{-1/2} p_{A1xz} +2\Psi _{xx} \Delta
^{-1/2} p_{A1xz} +   \\
2\Psi _{xz} \Delta^{-1/2} p_{A1zz} ] +  \Delta^{-1/2}
\left[ -\left( \Psi _{xz} \right) ^{2} +\Psi _{xx} \Psi_{zz} ) \right]+\\
     \epsilon \left[ -\Psi _{z} \Delta ^{-1} (\phi \Psi )_{xxx} + \Psi _{x}
\Delta ^{-1} (\phi \Psi )_{xxz} \right]   + \\
\epsilon \Delta ^{-1} \phi \Delta ^{-1} \partial ^{2} /\partial
x^{2} \left[ \Psi _{z} \Delta \Psi _{x} -\Psi _{x} \Delta \Psi
_{z} \right] +  O(\epsilon ^{2} )\
   \end{array}
\end{equation}
Between the nonlinear terms one can recognize interaction ($A1-TS
$ )and generation ones ($TS-TS$). The equation for   $p_{A2}$ is
obtained by projecting $P_{A2}$ and looks very similar. The
complete system includes this equation and (\ref{a}),(\ref{NTS}).
The system covers all possible processes description up to
quadratic terms approximation. Here a multimode TS waves in the
(OS) equation solutions basis could be incorporated . The
long-wave limit of such disturbance leads to coupled KdV system
\cite{LTSA}.

\section{Resonance interaction of acoustic and TS modes.}

Equations (\ref{NTS1}), (\ref{a}) form a coupled system of
evolution equations for interacting acoustic and TS modes. In the
case of the TS mode generation by an incoming first acoustic mode,
the early stage of evolution (for small amplitudes of TS mode) is
defined by a system:
\begin{equation}\label{TS2}
\begin{array}{c}
   \Delta \Psi _{t} +V_{0} \Delta \Psi _{x} -\Psi _{x} \cdot \phi _{z} -%
{Re} ^{-1} \Delta ^{2} \Psi =\\
  -\epsilon \left( \Delta \Psi \cdot \Delta ^{1/2} p_{A1} +\Delta
\Psi_{x} \cdot\Delta ^{-1/2} p_{A1x} +\Delta \Psi_{z} \cdot
\Delta^{-1/2} p_{A1z} \right)
\end{array}
\end{equation}
\begin{equation}\label{a1}
  \begin{array}{c}
    \epsilon \left( \partial p_{A1} /\partial t+V_{0} \partial p_{A1} /\partial
x\right) +\Delta ^{1/2} p_{A1} =
\frac{\epsilon }{2} \left[-\Psi _{z} p_{A1x} +\Psi _{x} p_{A1z} \right]  + \\
\epsilon \Delta ^{-1/2} [ -\Psi _{z} \Delta ^{1/2} p_{A1x} +\Psi
_{x} \Delta ^{1/2} p_{A1z} -2\Psi _{xz} \Delta ^{-1/2}
p_{A1xx} -\\
2\Psi _{zz} \Delta ^{-1/2} p_{A1xz} +2\Psi _{xx} \Delta ^{-1/2}
p_{A1xz} +2\Psi _{xz} \Delta ^{-1/2} p_{A1zz}].] \
  \end{array}
\end{equation}
All quadratic terms relating to $TS-TS$ interaction, are left out
of account . For simplicity, we consider only the first incoming
acoustic mode.

As it follows from the discussion in the introduction, let us find
a  solution in the form:
\begin{equation}\label{anza}
 p_{A1} (x,z,t)=A_{1} (\mu x,\mu t)\pi _{1} \exp (i(\omega _{1}
t-k_{1} x))+A_{2} (\mu x,\mu t)\pi _{2} \exp (i(\omega _{2}
t-k_{2} x))+c.c.
\end{equation}
\begin{equation}\label{anzTS}
 \Psi (x,z,t)=B_{3} (\mu x,\mu t)\psi _{3} (z)\exp (i(\omega _{3}
t-k_{3} x))+B_{4} (\mu x,\mu t)\psi_{4} (z)\exp (i(\omega_{4} t -
k_{4} x))+ c.c.
\end{equation}
where $\Pi _{1} =\pi_{1} (k_{1} ,\omega_1 ,z)\exp (i(\omega _{1}
t-k_{1} x))$ , $\Pi _{2} =\pi _{2} (k_{2} ,\omega_2 ,z)\exp
(i(\omega _{2} t-k_{2} x))$ are planar waves. $\Pi _{1} $
satisfies the linear evolution equation (\ref{A1})
\begin{equation}\label{a1}
 \partial \Pi _{1}  /\partial t + V_{0} \partial \Pi _{1} /\partial x+\epsilon
^{-1} \Delta ^{1/2} \Pi _{1} = 0.
\end{equation}
that leads to the equivalent equation for  $\Pi _{1}$
 \begin{equation}\label{pi1}
i\omega _{1}  \Pi _{1}  - ik_1 V_{0} \Pi _{1} +\epsilon ^{-1}
\Delta ^{1/2} \Pi _{1} = 0.
\end{equation}
Suppose the vertical gradients of all wave functions inside the
viscous layer are much bigger then horizontal ones. From
experiments (e.g. \cite{Koz}), it is known, that wavelength of TS
mode is much greater than a thickness of the boundary viscous
layer for common values of Reynolds number. So, the operator
$\Delta ^{1/2} $  may be evaluated as the generalized operator
(Taylor) series with respect to $\partial_z/k_1$. Hence the
operator radical in the first approximation is evaluated via
Gataux derivative as
 $$
 \Delta ^{1/2} \Pi _{1}=  \sqrt[2]{-k_1^2 + \partial_z^2}\Pi _{1}=
  \imath k_1   \sqrt[2]{1 - \partial_z^2/k_1^2} \Pi _{1} \approx  \imath
  k_1(1 - \partial_z^2/2k_1^2)\Pi _{1}
 $$
and, for (\ref{pi1}), one arrives to the ordinary differential
equation, that we can consider as a spectral problem with the
spectral parameter $k_1$,
\begin{equation}\label{pi1app}
 (1 - \frac{k_1}{\omega_{1}} V_{0}) \pi _{1} + \frac{k_1}{\omega _{1}
 \epsilon}(1 - \frac{1}{2k_1^2}\partial_z^2)\pi _{1}  = 0.
\end{equation}
The same equation obviously define $\pi_2$, it is enough to change
indices $1\rightarrow 2$ in the operator.
The functions $\psi _{3}(z)$ , $\psi _{4} (z)$ are solutions of
the OS equation (\ref{12}) suitable for a concrete problem. $A_{1}
$ ,.., $B_{4} $ are slowly varying functions of x, t, that's why
an additional small parameter $\mu$ is introduced , generally,
they are complex functions.
Calculating the right-hand nonlinear expressions, we take only
first term in series to avoid small terms of the higher order.

Let us discuss a possibility of four-waves resonance. Examining
the algebraic relations between parameters yields the appropriate
conditions:
\begin{equation}\label{res}
 \omega _{1} =\omega _{2} -\omega _{3} ,  \omega _{2} = \omega
_{1} +\omega _{3} ,  \omega _{3} = \omega _{1} + \omega _{4},
 \omega _{4} = \omega _{3} -\omega _{1}
\end{equation}

Substituting the formulas (\ref{anza}, \ref{anzTS}) to
(\ref{a1},\ref{TS2}), and picking up the resonant terms only, one
goes to the further system of equations (complex conjugate values
marked with asterisks, $ k_{1} - k_{2} + k_{3} =\Delta k, k_{3}
-k_{1} - k_{4} = \Delta k')$:

$\mu \left( \epsilon A_{1T}  \pi _{1} +A_{1X} \left( \epsilon
V_{0} \pi _{1} -ik_{1} \int\limits_{0}^{z}\pi _{1} dz \right)
\right) =$

$\epsilon A_{2} B_{3}^{*} \left( 1.5\ (ik_{3}^{*} \psi _{3}^{*}
\pi _{2z} +ik_{2} \psi _{3z}^{*} \pi _{2} )\right)$

+$\int\limits_{0}^{z}\left( ik_{3}^{*} \psi _{3z}^{*} \pi _{2z}
+ik_{2} \psi _{3zz}^{*} \pi _{2} \right) dz  e^{i( \Delta k )x} $,

$\mu \left( \epsilon A_{2T} \pi _{2} +A_{2X} \left( \epsilon V_{0}
\pi _{2} -ik_{2} \int\limits_{0}^{z}\pi _{2} dz \right) \right) =$

$\epsilon A_{1} B_{3}^{*} \left(1.5\ ( -ik_{3} \psi _{3}^{*} \pi
_{1z} +ik_{1} \psi _{3z}^{} \pi _{1} )\right)$ +

$ \int\limits_{0}^{z}\left( -ik_{3} \psi _{3z}^{*} \pi _{1z}
+ik_{1} \psi _{3zz}^{*} \pi _{1} \right) dz  e^{i(- \Delta k )x}
$,

$\mu \left( B_{3T} \psi _{3zz} +B_{3X} \left[ \left( V_{0}
+4ik_{3} Re^{-1} \right) \psi _{3zz} +\left( 2k_{3} \omega _{3}
-\phi _{z} \right) \psi _{3} \right] \right) =$

$-\epsilon A_{1} B_{4}^{} \left( \psi _{4zz}^{} \pi _{1z}
-k_{1}^{} k_{4} \psi _{4zz}^{} \int\limits_{0}^{z}\pi _{1} dz
+\psi _{4zzz}^{} \pi _{1} \right) e^{i( \Delta k' )x} $,

$\mu \left( B_{4T} \psi _{4zz} +B_{4X} \left[ \left( V_{0}
+4ik_{4} Re^{-1} \right) \psi _{4zz} +\left( 2k_{4} \omega _{4}
-\phi _{z} \right) \psi _{4} \right] \right) =$

$-\epsilon A_{1}^{*} B_{3}^{*} \left( \psi _{3zz}^{*} \pi
_{1z}^{*} +k_{1}^{*} k_{3} \psi _{3zz}^{*} \int\limits_{0}^{z}\pi
_{1}^{*} dz +\psi _{3zzz}^{*} \pi _{1}^{*} \right) e^{i( - \Delta
k' )x} $

We take into account that acoustic wavenumbers are real, but
wavenumbers of both T-S modes may be complex in the general case,
namely the points of real values form the neutral curve
\cite{Sch}.

This resulting equation may be considered as 4-wave resonance
equation but without synchronism condition \cite{LN}. The
coefficients of the equation depend on z, it is due to our choice
of the only one transverse mode for each "horizontal" one.
Following the lines of \cite{L}, that have a connection of
Galerkin numerical method,

 We continue the projecting procedure
considering the transverse modes as a basis. In such problems we
naturally arrive at two bases. One arises from TSW theory and its
origin is from OS equation (denoted by ). Other is from sound
problem. Of course, such bases are not orthogonal. Hence we could
only multiply each equation by the its own basic vector and
integrate across the boundary layer. The result is written in
"back"-re-scaled  variables: we put $\mu = 1$.
 \begin{equation}\label{4wave}
 \begin{array}{c}
    A_{1T}  + c_{a1} A_{1X} = n_{a1} A_{2} B_{3}^{*}  e^{i(\Delta k)x},\\
  A_{2T}  + c_{a2} A_{2X} = n_{a2} A_{1} B_{3}^{*}  e^{i(-\Delta k)x},\\
    B_{3T}  + c_{TS1}B_{3X}  =
-\epsilon n_{TS1} A_{1} B_{4}^{}  e^{i( \Delta k' )x},\\
  B_{4T} + c_{TS2}B_{4X}  =
-\epsilon n_{TS1} A_{1}^{*} B_{3}^{} e^{i( - \Delta k' )x}\\
 \end{array}
\end{equation}
 where the group velocities and nonlinear constants are expressed
 via the integrals across the boundary layer with a width
 $\delta$:
 \begin{equation}\label{const}
 \begin{array}{c}
  c_{a1} = \frac{\int\limits_{0}^{\delta} [\pi _{1}^2 - \epsilon^{-1}ik_{1}\pi _{1} \int_o^\delta\pi _{1} dz]dz' }{\int_0^\delta V_{0}\pi _{1}^2dz}  \\
 n_{a1} =  \frac{\int_0^\delta [ 1.5 ( -ik_{3}^{*} \psi
_{3}^{*} \pi _{2z} +ik_{2} \psi _{3z}^{*} \pi _{2}  t) +
\int\limits_{0}^{z'} \pi _{1} ( -ik_{3}^{*} \psi_{3z}^{*}\pi_{2z}
+ik_{2} \psi_{3zz}^{*} \pi _{2}  ) dz ]dz' }{\int_0^\delta V_{0}\pi _{1}^2dz}\\
  c_{a2} = \frac{\int\limits_{0}^{\delta}[ \pi _{2}^2 - \epsilon^{-1}ik_{2}\pi _{2} \int_0^\delta\pi _{2} dz]dz' }{\int_0^\delta V_{0}\pi _{1}^2dz}  \\
 n_{a2} =  \frac{\int_0^\delta  \pi _{2} [( 1.5 ( -ik_{3} \psi
_{3} \pi _{1z} +ik_{1} \psi _{3z} \pi _{1}  ) +
\int\limits_{0}^{z'} ( -ik_{3}  \psi _{3z}  \pi _{1z} +ik_{1}
\psi_{3zz}  \pi _{1}  ) dz )]dz'}{\int_0^\delta V_{0}\pi
_{1}^2dz}\\ c_{TS1} = \frac{\int_0^\delta \psi_3  [ ( V_{0} +i
4k_{3}Re^{-1}
 ) \psi _{3zz} +  ( 2k_{3} \omega _{3}
+\varphi_{z}  ) \psi _{3}  ]  )dz}{\int_0^\delta \psi_3 \psi _{3zz}dz } \\
n_{TS1} = \frac{\int_0^\delta \psi_3 ( \psi _{4zz}^{} \pi_{1z}
-k_{1}^{} k_{4} \psi _{4zz}^{} \int\limits_{0}^{z'}\pi _{1}
dz +\psi_{4zzz}^{} \pi _{1}  )dz'}{\int_0^\delta \psi_3 \psi_{3zz}dz}\\
c_{TS2} =  \frac{\int_0^\delta \psi_4  [ ( V_{0} +i4k_{4} Re^{-1}
 ) \psi _{4zz} +  ( 2k_{4} \omega _{4} +\varphi_{z}
 ) \psi _{4}  ]  )dz}{\int_0^\delta \psi_4 \psi_{4zz}dz }\\
 n_{TS2} = \frac{\int_0^\delta \psi_4 ( \psi _{3zz}^{} \pi_{1z}^* - k_{1}^{} k_{3} \psi _{3zz}^{} \int\limits_{0}^{z'}\pi_{1}^*dz +\psi_{3zzz}^{} \pi _{1}^* )dz'}{\int_0^\delta \psi_4 \psi_{4zz}dz}\\
 \end{array}
\end{equation}
 A structure of the obtained equations is the particular case of
 general N-wave system , that may be solved by special technics
valid for integrable equations \cite{LN}. The 4-wave approximation
may give rise to such solutions that exhibit effective energy
exchange between modes. The form of the nonlinearity is typical
for a 3-wave systems and even small fluctuations of a TS field
could initiate a rapid growth of both components if the acoustic
field is big enough. It is pure nonlinear instability that may be
supported by linear stability curve shift \cite{LPG}. The
numerical evaluation of the integrals in constants \ref{const}
need rather compulsory calculations.  The solution of the system
(\ref{4wave}) pose also the separate problem. We plan to present
the results in a next paper.
 \section{Conclusion}
  The resulting system (\ref{NTS}),(\ref{a}) and the equation of
the opposite directed A-mode could be considered as a basic one
for all-perturbations over a BL description. We also would note
that the boundary layer width may depend on x. A slow dependence
is usually accepted and do not change the general structure of the
expressions. As the resonant as a non-resonant processes may be
studied with the acoustic waves separated. The modes have the
different scales, hence a numerical modelling of the mutual
generation and control also could be more effective.

\end{document}